\documentclass[12pt]{iopart}
\usepackage{graphicx}
\begin{document}

\title{A Study on Shape of Te Isotopes in Mean Field Formalism}

\author{T. Bayram$^*$}

\address{Department of Physics, Sinop University, Sinop, Turkey}
\ead{t.bayram@ymail.com}

\author{A. H. Yilmaz}

\address{Department of Physics, Karadeniz Technical University, Trabzon, Turkey}

\begin{abstract}
The systematic investigation of ground-state shape evolution from
$\gamma-$unstable O(6) to spherical U(5) for even$-$even
$^{112-134}$Te has been presented by using the quadrupole moment
constrained Hartree-Fock-Bogoliubov (HFB) method with the Skyrme
force SLy4. $^{124}$Te has been pointed out as to be the possible
critical-point nucleus with E(5) symmetry.
\end{abstract}



\maketitle

\section{Introduction}
Spherical vibrator, rotational ellipsoid, and deformed shapes of
nuclei are related with modes of collective motion and geometric
shapes of nuclei~\cite{Bohr75,Ring80,Greiner96,Cwiok05}. When the
number of protons and neutrons is changed, nuclei can show changes
of their energy levels and electromagnetic transition rates among
collective modes of nuclei. Transitions from one kind of collective
behavior to another are named shape phase transition. These
transitions are quantum phase transitions (QPTs)~\cite{Iachello06}.
They are different from thermal phase transitions which occur as a
function of temperature. This implies changes in the shape of the
nucleus. The control parameter is the number of nucleons. In the
last decade, many researchers have given insights into the evolution
of structure of nuclei (in particular transitional regions of rapid
change)~\cite{Casten07}. They have used the concepts of QPTs, phase
coexistence and the critical-point symmetries
(CPS)~\cite{Iachello00,Iachello01} proposed by Iachello, as well as
a raft of geometrical
models~\cite{Caprio02,Iachello03,Fortunato04,Bonatsos04,Iachello05,Bonatsos06}.\\

Theoretically, QPTs have been mostly studied in the Interacting
Boson Model (IBM). It holds the U(5), SU(3) and O(6) symmetries
within the simplest U(6) symmetry. The U(5), SU(3) and O(6)
dynamical symmetries correspond to the shape phase of a spheroid,
axially prolate rotor and $\gamma-$soft rotor,
respectively~\cite{Iachello87}. Using the model, the authors pointed
out that second order shape phase transition occurs between U(5) and
O(6) while first order shape phase transition occurs between O(6)
and SU(3)~\cite{Ginocchio80,Dieperink80}. More recently, Iachello
has pointed out that the properties of nuclei lying at the
critical-point of a shape phase transition can be described by
special solutions of the Bohr Hamiltonian, called as critical-point
symmetries. The E(5) critical-point symmetry~\cite{Iachello00}
corresponds to the second order critical-point between U(5) and
O(6), while the X(5) critical-point symmetry~\cite{Iachello01}
corresponds to the first order transition between U(5) and SU(3).
Experimentally, The E(5) and X(5) symmetry has been realized in the
spectrum of $^{134}$Ba~\cite{Casten00} and
$^{152}$Sm~\cite{Casten01}, respectively. The introduction of
critical-point symmetries E(5) and X(5), and their experimental
realizations have triggered many works
on quantum phase transitions~\cite{Bonatsos07}.\\

In spite of the fact that the IBM and the solutions of Bohr
Hamiltonian are useful for determining shape phase transitions in
nuclei, mean field formalisms (e.g., HFB
method~\cite{Ring80,Flocard73,Decharge80} and relativistic mean
field (RMF) model~\cite{Serot86,Gambhir90,Ring96,Meng06}) which
provide a correct prediction of many nuclear phenomena have been
successfully used to study the shape phase transition in nuclei. The
RMF theory has been used to investigate the critical-point nuclei in
even$-$even Sm~\cite{Meng05} and Ce~\cite{Yu06} isotopes. In these
studies, $^{148,150,152}$Sm and $^{128,130,132,134}$Ce have been
suggested as an example of the possible critical-point nuclei with
X(5) symmetry. Beside, Ti isotopes have been examined in the HFB
method~\cite{Bayram12} and RMF model~\cite{Guo08} to investigate the
critical-point nuclei. In these studies, $^{48,52,60}$Ti and
$^{46,52,60}$Ti have been found as to be the possible candidate
critical-point nuclei with E(5) symmetry in the RMF model and HFB
method, respectively. A series of isotopes in rare-earth region have
been found as to be possible critical-point
nuclei~\cite{Fossion06,Rodriguez07}. Also, Mo isotopes were
investigated by using the RMF theory~\cite{Yao10,Yilmaz11} and
$^{94}$Mo has been suggested as to be $\gamma-$unstable nucleus in
Ref.~\cite{Yao10}. In these studies, potential energy curves (PECs)
come by quadrupole moment constrained calculations have been
examined in order to identifing of the critical-point nuclei.
Relatively flat PECs correspond to the critical-point nuclei with
E(5) symmetry, while PECs with a bump correspond to the X(5)
symmetry (The relation between shape phase transition and PECs can
be found in~\cite{Casten07,Eisenberg87}). However, for a
quantitative analysis of QPTs in nuclei electromagnetic transition
rates and ratios of excitation energies should be
calculated~\cite{Li10}. For this reason, the generator coordinate
method (GCM) has been used to perform configuration mixing of
angular-momentum and particle-number projected relativistic wave
functions restricted to axial symmetry in~\cite{Niksic07}. In recent
years, the GCM has been extended on triaxial
states~\cite{Bender08,Yao102,Rodriguez10}. Nevertheles, the
application of these methods in a systematic study of shape
transition is at present still very time-consuming because of its
triaxiality. It should be noted, however, that the potential energy
curves obtained from quadrupole moment constrained calculations are
important, and can provide a qualitative understanding of the shape
phase transition. Particularly, the evolution of the PECs along the
isotopic or
isotonic chains can be useful for investigation of shape phase transitions in nuclei.\\

Rapid shape changes in nuclei have been known for about half a
century~\cite{Casten07}. Classic shape transition regions take place
in the light Si-Mg region~\cite{Bromley57}, near $A=100$
($Z\sim40$)~\cite{Cheifetz70}, light rare-earth region ($A\sim150$)
and actinides. Beside, the $\gamma-$unstable character of nuclei in
the mass region $A=120-130$ was pointed out in~\cite{Gneuss71} many
years ago. And, $^{124}$Te in this region was experimentally
investigated and suggested to be a possible $\gamma-$soft
nucleus~\cite{Subber86}. In this study, constrained HFB method has
been used to obtain the ground-state properties of even$-$even
$^{112-134}$Te isotopes such as total binding energy and quadrupole
deformation. The shape evolution of Te isotopes has
been analized by examining their PECs.\\

\section{The theoretical framework and numerical details}
In the HFB formalism, a two-body Hamiltonian of a system of fermions
by means of a set of annihilation and creation operators ($c$,
$c^{\dagger}$) is given by
\begin{equation}
H=\sum_{n_{1}n_{2}}e_{n_{1}n_{2}}c_{n_{1}}^{\dagger}c_{n_{2}}+\frac{1}{4}\sum_{n_{1}n_{2}n_{3}n_{4}}\bar
\upsilon_{n_{1}n_{2}n_{3}n_{4}}c_{n_{1}}^{\dagger}c_{n_{2}}^{\dagger}c_{n_{4}}c_{n_{3}}\,,
\label{Hamiltonian}
\end{equation} where $\bar \upsilon_{n_{1}n_{2}n_{3}n_{4}}=\langle n_{1}n_{2}\mid V\mid
n_{3}n_{4}-n_{4}n_{3}\rangle$ are anti-symmetrized matrix elements
of the two-body N-N interaction. The ground-state wave function
$|\Phi\rangle$ is described as the quasi-particle vacuum
$\alpha_{k}|\Phi\rangle=0$ and the linear Bogoliubov transformation:
\begin{equation}
\alpha_{k}=\sum_{n}(U_{nk}^{*}c_{n}+V_{nk}^{*}c_{n}^\dagger),
\hspace{0.3 in}
\alpha_{k}^\dagger=\sum_{n}(V_{nk}c_{n}+U_{nk}c_{n}^\dagger)\,
\label{Bogoliubov}
\end{equation} provides connection between the quasiparticle operators
$(\alpha,\alpha^\dagger)$ and the original particle operators. The
basic building blocks of the theory are the density matrix and the
pairing tensor. In terms of the normal $\rho$ and pairing $\kappa$
one-body density matrices:
\begin{equation}
\rho_{nn'}=\langle\Phi|c_{n'}^\dagger
c_{n}|\Phi\rangle=(V^{*}V^{T})_{nn'}, \hspace{0.3 in}
\kappa_{nn'}=\langle\Phi|c_{n'}c_{n}|\Phi\rangle=(V^{*}U^{T})_{nn'}\,,
\label{matrices}
\end{equation} the expectation value of the
Hamiltonian (\ref{Hamiltonian}) could be expressed in terms of an
energy functional:
\begin{equation}
E[\rho,\kappa]=\frac{\langle\Phi|H|\Phi\rangle}{\langle\Phi|\Phi\rangle}=
\textrm{Tr}[(e+\frac{1}{2}\Gamma)\rho]-\frac{1}{2}\textrm{Tr}[\Delta\kappa^{*}]
\label{energyfunctional}
\end{equation} where
$\Gamma_{n_{1}n_{3}}=\sum_{n_{2}n{4}}\bar\upsilon_{n_{1}n_{2}n_{3}n_{4}}\rho_{n_{4}n_{2}}$
and
$\Delta_{n_{1}n_{2}}=\frac{1}{2}\sum_{n_{3}n{4}}\bar\upsilon_{n_{1}n_{2}n_{3}n_{4}}\kappa_{n_{3}n_{4}}$.
Modern energy functional (\ref{energyfunctional}) includes terms
that cannot be simply related to some prescribed effective
interaction~\cite{Bender03}. In terms of Skyrme forces, the HFB
energy (\ref{energyfunctional}) has the form of local energy density
functional:
\begin{equation}
E[\rho,\tilde{\rho}]=\int d^{3}\textrm{H}(\textbf{r}),
\label{skyrmeefunctional}
\end{equation} where
$\textrm{H}(\textbf{r})=H(\textbf{r})+\tilde{H}(\textbf{r})$ is the
sum of the mean field and pairing energy densities. The variation of
the energy (\ref{skyrmeefunctional}) according to the particle local
density $\rho$ and pairing local density $\tilde{\rho}$ results in
Skyrme HFB equations:

\begin{eqnarray}
\sum_{\sigma'}\left(\begin{array}{cc} h(\textbf{r},\sigma,\sigma') & \tilde{h}(\textbf{r},\sigma,\sigma') \\
\tilde{h}(\textbf{r},\sigma,\sigma') & -h(\textbf{r},\sigma,\sigma')
\end{array} \right)\left(\begin{array}{c} U(E,\textbf{r}\sigma') \\
V(E,\textbf{r}\sigma')\end{array} \right) \nonumber \\=\left(\begin{array}{cc} E+\lambda & 0\\
0 & E-\lambda)\end{array}\right)\left(\begin{array}{c} U(E,\textbf{r}\sigma) \\
V(E,\textbf{r}\sigma) \end{array}\right), \, \label{shfb}
\end{eqnarray} where $\lambda$ is chemical potential. Local fields $h(\textbf{r},\sigma,\sigma')$
and $\tilde{h}(\textbf{r},\sigma,\sigma')$ can be calculated in the
coordinate space~\cite{Ring80,Stoitsov05}.\\

The HFB equations~(\ref{skyrmeefunctional}) have been solved by
expanding quasi-particle wave functions that conserve axial symmetry
and parity on a harmonic oscillator basis expressed in coordinate
space prescribed by Stoitsov \textit{et al}.~\cite{Stoitsov05} in
the present study. For pairing, the zero-range pairing interaction
is taken into account and Lipkin-Nogami method is employed (Further
details are given in Ref.~\cite{Stoitsov05}). The oscillator
parameter $b_{0}$ is chosen as to be
$b_{0}=\sqrt{2(\hbar^{2}/2m)(49.2A^{-1/3})}$. In order to obtain the
potential energy curves (PECs) in the present study, the standard
quadratic form of the quadrupole
constraint~\cite{Flocard73,Stoitsov05} has been performed. The
standard quadratic form can be interpreted by the formula
$E^{Q}=C_{Q}(\langle \hat{Q}\rangle-\bar{Q})^{2}$, where $C_{Q}$ is
the stiffness constant, $\langle \hat{Q}\rangle$ is the average
value of the mass quadrupole moment operator
($\hat{Q}=2z^{2}-r^{2}$) and $\bar{Q}$ is the constraint value of
the quadrupole moment. For describing the deformation of nuclei, the
quadrupole deformation parameter is commonly used rather than
quadrupole moments. The relation between the constraint quadrupole
moment $\bar{Q}$ and constraint quadrupole deformation parameter
$\bar{\beta_{2}}$ is given by the formula
$\bar{\beta_{2}}=\sqrt{\pi/5}\langle \bar{Q} \rangle/ \langle r^{2}
\rangle$~\cite{Stoitsov05}. 16 oscillator shells have been taken
into account in the present calculations. There can be found a
number of effective Skyrme forces in literature for correct
prediction of the nuclear ground-state properties of
nuclei~\cite{Bartel82,Baran95,Chabanat98}. In this work, the Skyrme
force SLy4~\cite{Chabanat98} has been used to calculate ground-state
properties of even$-$even Te isotopes with 60 $\leq N \leq$ 82.

\section{Results and discussions}

The calculated total binding energies for even$-$even $^{112-134}$Te
isotopes obtained from the constrained HFB method with the SLy4
parameter set are tabulated in Table~\ref{tab1}. Also, the
predictions of RMF model~\cite{Lalazissis99} and the experimental
data~\cite{Audi03} are listed for comparison. The total binding
energies of all isotopes are reproduced well by the SLy4 Skyrme
force. The deviations are at most 0.3\%. Also, as can be seen in
Table~\ref{tab1}, the predictions of RMF model with NL3 interaction
are in a good agreement with the experimental data. The mean
differences between experimental data and the predictions of the HFB
method and RMF model with NL3 interaction are 2.532 and 1.688 MeV,
respectively. For this reason, it can be pointed out that the
predictions of both HFB method and RMF model are good in
describing the ground-state binding energies of $^{112-134}$Te.\\

\begin{figure}
\begin{center}
\includegraphics[width=0.6\textwidth]{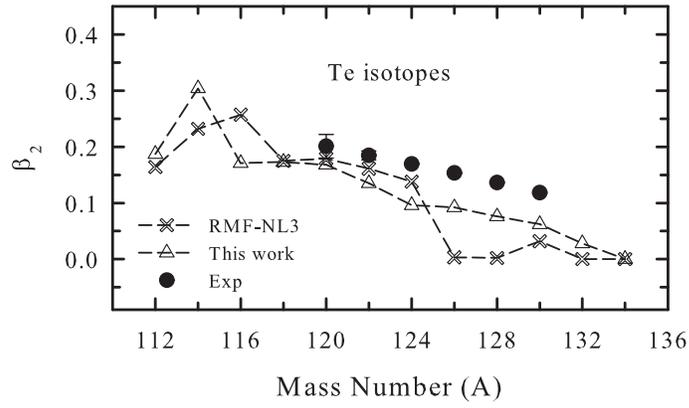}
\caption{The ground-state quadrupole deformation parameter for Te
isotopes. The predictions of the HFB method with SLy4 Skyrme force
are compared with the those of the RMF model with NL3 interaction
and experimental results.\label{fig1}}
\end{center}
\end{figure}

The mean field formalism based on the Hartree-Fock approximation
with phenomenological effective interactionsis is important in the
microscopic description of nuclei~\cite{Vautherin72,Gogny75}. It
allows a unified description of the ground-state properties for
nuclei throughout the nucleidic chart. One of the great achievement
of the theory is that not only it can reproduces binding energies
and densities, but it also provides a good description of the size
of the ground-state deformations in nuclei~\cite{Vautherin73}. In
this study, the quadrupole deformation parameters $\beta_{2}$ for
$^{112-134}$Te have been obtained from constrained HFB method with
SLy4 Skyrme force. They are shown in Fig.~\ref{fig1}. Also, the
predictions of RMF model with NL3 interaction~\cite{Lalazissis99}
and the experimental data~\cite{Raman01} are shown for comparison.
As can be seen in the Fig.~\ref{fig1}, the calculated $\beta_{2}$
values obtained from HFB method with SLy4 parameters are in good
agreement with the experimental data. Only amplitude of quadrupole
deformation parameter $\beta_{2}$ values obtained from both of the
HFB method and RMF model are given in the Fig.~\ref{fig1} and the
exact values of $\beta_{2}$ are listed in Table~\ref{tab2}. It
should be noted, however, that $\beta_{2}$ cannot be observed
directly from an experiment. To obtain the $\beta_{2}$ value from an
experiment, a conventional way is that the electric quadrupole
transition rate from ground-state $0^{+}$ to the $2^{+}$ state
$B(E2)\uparrow$ can be used~\cite{Raman01}. The correlation between
$B(E2)\uparrow$ and $\beta_{2}$ is given by the formula
$\beta_{2}=(4\pi/3ZR_{0}^{2})[B(E2)\uparrow/e^{2}]^{1/2}\,$ where
$R_{0}=1.2A^{1/3}$. The formula based on rigid rotor cannot always
represent a parameter of deformation. The extracting of $\beta_{2}$
is questionable in the case of spherical nuclei, because
$B(E2)\uparrow$ connects vibrational states in the spherical nuclei.
In particular, the radius $R_{0}$ is so small for light nuclei. This
elicites a very large $\beta_{2}$ deformation with the formula.
However, in medium-mass and heavy region usage of the formula is
siutable~\cite{Guo08}.\\

\begin{figure}
\begin{center}
\includegraphics[width=1.0\textwidth]{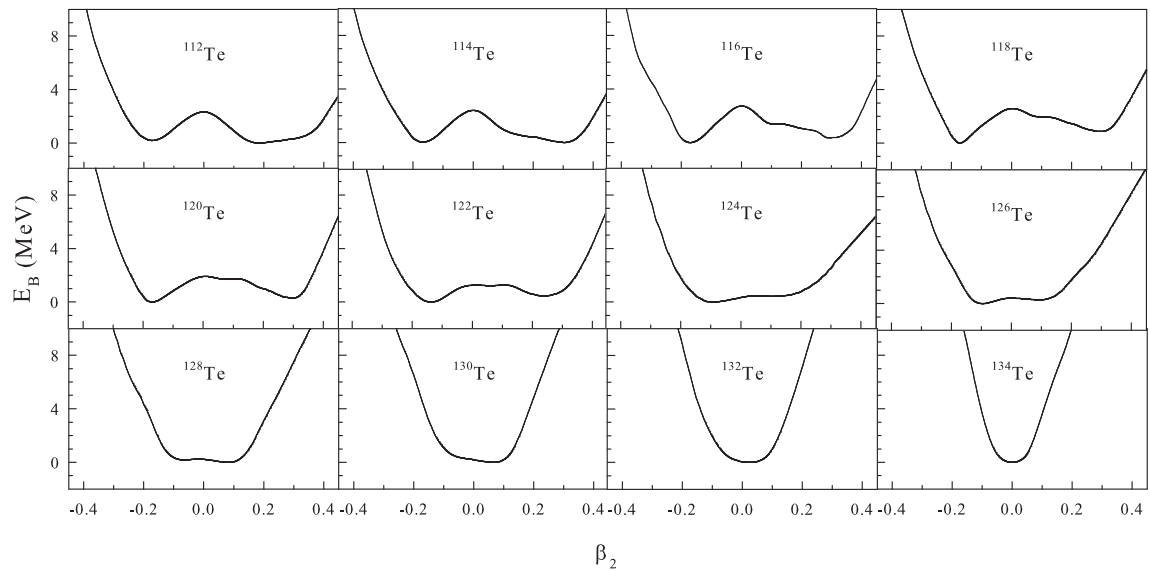}
\caption{The potential energy curves for $^{112-134}$Te obtained
from the constrained HFB method with SLy4 Skyrme force.\label{fig2}}
\end{center}
\end{figure}

In this study, the constrained HFB method with Skyrme force SLy4 is
employed for the investigation of shape evolution of even$-$even
$^{112-134}$Te because of its success in describing the binding
energies and quadrupole deformation parameters for Te isotopes. In
Fig.~\ref{fig2}, the potential energy curves (PECs) for
$^{112-134}$Te. In the figure, the total binding energy of Te
isotopes for the ground-state is taken as to be the reference. In
the Fig.~\ref{fig2}, starting from $^{112}$Te to $^{120}$Te, the
nuclei have oblate shape. In the PECs of the $^{122-126}$Te, their
barriers against deformation are weak which means that these nuclei
may be in a transitional region. In particular, the PEC of
$^{124}$Te in the Fig.~\ref{fig2} seems flat from $\beta_{2}=-0.2$
to $\beta_{2}=0.25$. Through these $\beta_{2}$ ranges, the variation
of the binding energies in the PEC of $^{124}$Te are less than 0.4
MeV. This implies that the barriers against deformation are very
weak, and $^{124}$Te maybe a possible example of the critical-point
nuclei with E(5) symmetry. With increasing of neutron numbers
starting from the $^{128}$Te to $^{132}$Te, shape of Te isotopes
have become prolate and finally $^{134}$Te which has shell closure
with magic neutron numbers $N=82$ are found to be as spherical.\\

In Table~\ref{tab3}, the differences of the binding energy between
the spherical-state and the ground-state of even-even $^{112-134}$Te
isotopes are shown to understanding of how the shape of the Te
isotopes changes with the neutron number as an additional evidence
to the results of the PECs. They can show how soft the nucleus is
against deformation. The calculated binding energy differences
between the spherical-state and the ground-state of $^{112-134}$Te
isotopes changes from 0 to 2.572 MeV. Drastic changes are clearly
visible around $^{120-124}$Te in the binding energy differences.
There is a clear jump appearing at $^{124}$Te which implies that
$^{124}$Te can be a possible candidate critical-point nuclei with
E(5) symmetry.\\

As an additional evidence for confirmation of the result of this
study, the ratios of experimental excitation energies of $^{124}$Te
nucleus~\cite{Subber86} are given in Table~\ref{tab4}. Also, the
U(5), X(5), SU(3), E(5) and O(6) symmetry predictions are listed for
comparison~\cite{Iachello00,Iachello01,Subber86}. The characteristic
ratio $R_{4/2}=E(4_{1}^{+})/E(2_{1}^{+})$ and the ratio of the
energies of the first two excited 0$^{+}$ states
$R_{0/2}=E(0_{2}^{+})/E(2_{1}^{+})$ are tabulated. As can be seen in
Table~\ref{tab4}, the E(5) symmetry values obtained from solution of
Bohr-Mottelson differential equations for $R_{4/2}$ and $R_{0/2}$ is
2.20 and 3.03, respectively. They are closer to the observed ratios
$R_{4/2}=2.07$ and $R_{0/2}=2.75$ which means that $^{124}$Te may
hold the E(5) symmetry.

\section{Summary}
The total binding energies and quadrupole deformation parameters for
even$-$even $^{112-134}$Te isotopes have been calculated in the
constrained HFB method with Skyrme SLy4 force as in a good agreement
with experimental data. The ground-state shape evolution of these
nuclei are investigated by using the potential energy curves.
$^{124}$Te has been found as to be an example for possible
critical-point nucleus, which marks the phase transition between
spherical U(5) and $\gamma-$unstable shapes O(6). It should be
noted, however, that this work investigates only the PECs of Te
isotopes with respect to the $\beta$ degree of freedom. For a
quantitative identifying of the E(5) symmetry in Te isotopes, one
should go betond mean field.


\newpage

\begin{table}
\begin{center}
\caption{The total binding energy for the ground-state of
$^{112-134}$Te in units of MeV.\label{tab1}}
\begin{tabular}{lccc}
\hline
      & This work & RMF~\cite{Lalazissis99} & Exp~\cite{Audi03}\\
\hline \hline
$^{112}$Te & $937.821$ & $938.880$ & $940.610$ \\
$^{114}$Te & $958.058$ & $959.060$ & $961.337$ \\
$^{116}$Te & $977.630$ & $978.530$ & $980.860$ \\
$^{118}$Te & $996.680$ & $997.600$ & $999.454$ \\
$^{120}$Te & $1014.148$ & $1015.640$ & $1017.281$ \\
$^{122}$Te & $1030.893$ & $1032.530$ & $1034.333$ \\
$^{124}$Te & $1047.472$ & $1049.160$ & $1050.686$ \\
$^{126}$Te & $1063.296$ & $1066.980$ & $1066.368$ \\
$^{128}$Te & $1078.858$ & $1080.750$ & $1081.439$ \\
$^{130}$Te & $1094.204$ & $1096.430$ & $1095.941$ \\
$^{132}$Te & $1108.850$ & $1112.220$ & $1109.914$ \\
$^{134}$Te & $1123.508$ & $1126.430$ & $1123.435$ \\
\hline
\end{tabular}
\end{center}
\end{table}

\clearpage

\begin{table}
\begin{center}
\caption{The ground-state quadrupole deformation parameter
$\beta_{2}$ for Te isotopes.\label{tab2}}
\begin{tabular}{lccc}
\hline
      & This work & RMF~\cite{Lalazissis99} & Exp~\cite{Audi03}\\

\hline \hline
$^{112}$Te & $0.187$ & $0.164$ &  \\
$^{114}$Te & $0.304$ & $0.232$ &  \\
$^{116}$Te & $-0.171$ & $0.257$ &  \\
$^{118}$Te & $-0.173$ & $0.175$ &  \\
$^{120}$Te & $-0.169$ & $0.179$ & $0.201$ \\
$^{122}$Te & $-0.135$ & $0.161$ & $0.185$ \\
$^{124}$Te & $-0.096$ & $0.138$ & $0.170$ \\
$^{126}$Te & $-0.093$ & $-0.003$ & $0.153$ \\
$^{128}$Te & $0.076$ & $-0.002$ & $0.136$ \\
$^{130}$Te & $0.062$ & $0.032$ & $0.118$ \\
$^{132}$Te & $0.028$ & $0.000$ &  \\
$^{134}$Te & $-0.005$ & $0.000$ &  \\
\hline
\end{tabular}
\end{center}
\end{table}

\clearpage

\begin{table}
\begin{center}
\caption{The difference of the total binding energy (in units of
MeV) between the spherical state and the ground-state of
$^{112-134}$Te obtained by the constrained HFB method with SLy4
Skyrme force.\label{tab3}}
\begin{tabular}{lc}
\hline
Nuclei      & HFB-SLy4 \\
\hline \hline
$^{112}$Te & $2.329$ \\
$^{114}$Te & $2.397$ \\
$^{116}$Te & $2.519$ \\
$^{118}$Te & $2.572$ \\
$^{120}$Te & $1.900$ \\
$^{122}$Te & $1.259$ \\
$^{124}$Te & $0.354$ \\
$^{126}$Te & $0.410$ \\
$^{128}$Te & $0.227$ \\
$^{130}$Te & $0.160$ \\
$^{132}$Te & $0.053$ \\
$^{134}$Te & $0.000$ \\
\hline
\end{tabular}
\end{center}
\end{table}

\clearpage

\begin{table}
\begin{center}
\caption{The ratios of available experimental excitation energies
for $^{124}$Te isotopes with some theoretical predictions for
comparison~[6$-$8, 44].
\label{tab4}}
\begin{tabular}{lccc}
\hline
     & $R_{4/2}$ & $R_{0/2}$ \\
\hline \hline
U(5) & $2.00$ & $2.00$ \\
X(5) & $2.91$ & $5.67$ \\
SU(3)& $3.33$ & $\gg2$ \\
E(5) & $2.20$ & $3.03$ \\
O(6) & $2.50$ & $4.50$ \\
\hline
Exp  & $2.07$ & $2.75$ \\
\hline
\end{tabular}
\end{center}
\end{table}

\clearpage

\clearpage

\clearpage

\end{document}